\documentstyle[preprint,prb,aps]{revtex}
\newcommand{\w}{\omega } 
\newcommand{\psil}{\psi_{\lambda} } 
\newcommand{\psilpsil}{\langle \psi_{\lambda}| \psi_{\lambda}\rangle } 
\begin{document}
\draft 
\title{Multifractality of the quantum Hall wave functions\\
in higher Landau levels} 
\author{Takamichi Terao and Tsuneyoshi Nakayama} 
\address{Department of Applied Physics, Hokkaido University, 
Sapporo 060, Japan} 
\author{Hideo Aoki}
\address{\it Department of Physics, University of Tokyo, 
             Hongo, Tokyo 113, Japan} 
\date{\today}

\maketitle

\begin{abstract}
To probe the universality class 
of the quantum Hall system at the metal-insulator critical point, 
the multifractality of the wave function $\psi$ is studied 
for higher Landau levels, $N=1,2$, for various range $(\sigma )$ 
of random potential.  
We have found that, while the 
multifractal spectrum 
$f(\alpha)$ (and consequently the 
fractal dimension) does vary with $N$, 
the parabolic form for 
$f(\alpha)$ indicative of a log-normal 
distribution of $\psi$ persists in 
higher Landau levels.  
If we relate the multifractality with the scaling of localization via 
the conformal theory, 
an asymptotic recovery 
of the single-parameter scaling with 
increasing $\sigma$ is seen, 
in agreement with Huckestein's irrelevant scaling field argument.  
\end{abstract}

\vspace*{0.5cm}

\pacs{73.40Hm, 64.60Ak, 71.30+h} 

\newpage 

It is an intriguing question to ask whether the transition to the 
Anderson-localized regime in disordered systems is regarded as 
a phase transition.  
A suggestion came in 1983 to the effect that 
the wave function at the Anderson transition should be 
self-similar (fractal) on all length scales 
just as the critical point in a phase transition 
is characterized by a diverging correlation 
length with the states being scale-invariant.\cite{aoki83}  
To be more precise, the self-similarity in critical wave functions 
extends beyond a single scale transformation, 
so the idea has been subsequently developed into 
the multifractal analysis.\cite{castellani,schreiber,huckerev}

The criticality of the wave function proposed in general has 
been analyzed for the quantum Hall (QH) system in 
particular,\cite{aoki83} where the delocalized states, 
marginally allowed to appear in two dimensions$(d=2)$ in the presence of 
magnetic fields, coalesce into a single point (the center of each Landau 
level) on the energy axis.  
The multifractal analysis has in fact proved to be essential for a 
full description of the criticality in the QH system 
as recently reviewed by Huckestein.\cite{huckerev}  

On the other hand, the scaling of localization in QH systems is 
still some way from a full understanding: 
a finite-size scaling study,\cite{AA}  
which looks into the scaling of $\xi_M/M =f(M/\xi )$ 
where $\xi_M$ is the 
localization length in long strips of width $M$, 
indicates 
that the single-parameter scaling 
holds for the QH system for 
the Landau index $N=0$, while a two-parameter scaling seems to be 
required for higher ($N\geq 1$) ones.  
In an attempt to reconcile the apparent lack of universality, 
Huckestein\cite{huck}   
has suggested that, if one introduces an irrelevant scaling 
length $\xi_{\rm irr}$ to describe 
$\xi_M/M$ as a function of two variables, $(M/\xi )$ and 
$(M/\xi_{\rm irr})$,\cite{huckerev} 
the single-parameter scaling is 
recovered asymptotically on length scales $> \xi_{\rm irr}$.  
The meaning of the second length scale is not fully 
clarified\cite{evers} except 
that the existence of two parameters is reminiscent of Pruisken's 
two-parameter scaling that involves the Hall ($\sigma_{xy}$) 
and longitudinal ($\sigma_{xx}$) conductivities.  

Here natural questions are: (i) how the multifractality will 
depend on the Landau index, and (ii) can this possibly related to 
the scaling, which are exactly the purpose of the present paper. 
It has in fact long been recognized\cite{AA,Ando84} that 
the localization length $\xi$ wildly grows with 
the Landau index $N$, 
where a mapping onto the 
nonlinear $\sigma$ model gives an estimate\cite{huckerev} 
$\xi \sim \ell{\rm exp}(\sigma_{\rm SCBA}^2)$ 
with $\sigma_{\rm SCBA} \propto (N+1/2)$ being $\sigma_{xx}$ 
in the self-consistent Born approximation\cite{Ando} and $\ell$ 
the magnetic length.   
The $N$ dependence indicates that 
the scaling has to be discussed on much larger length scales 
for higher levels, which is indeed one 
motivation for introducing the irrelevant length.  
There, we can control the situation if the irrelevant 
scaling idea is valid: 
the irrelevant length scale $\xi_{\rm irr}$ 
is shown\cite{huck} to rapidly decrease 
when the correlation length 
of the random potential $\sigma$ 
is increased from $\ll \ell$ to $\sim \ell$. 

Given this background, it will be 
intriguing how the multifractality will vary with 
$N$ and/or $\sigma$.   
We can also propose that the multifractal analysis provides a way to 
probe the scaling if we combine the multifractality and the 
scaling.  
Jan\ss en\cite{Janssen} has actually suggested that the scaling amplitude 
$\Lambda_c = {\rm lim}_{M\rightarrow \infty}\xi_M/M$ 
should be related with 
the multifractal spectrum $f(\alpha)$ via 
$\Lambda_c = [\pi (\alpha_0-d)]^{-1}$, 
where $\alpha_0$ is the 
position of the peak in $f(\alpha)$ against $\alpha$ and 
$d=2$ is the spatial dimension of the system.    
This is obtained by 
relating the typical finite-size scaling variable 
with the multifractal behavior of the critical wave function 
through the conformal mapping.\cite{Cardy,Pichard}

The Hamiltonian of the QH system is given by 
\begin{equation} 
  {\cal H}=\sum_{NX} |NX\rangle \left( N+{1\over 2} \right) 
   \hbar \w_c  \langle NX| \nonumber \\
   + \sum_{NX} \sum_{N'X'}    |NX\rangle \langle NX| 
     V |N'X'\rangle \langle N'X'| \ 
\end{equation} 
with the Landau wave function $|NX\rangle$, 
the cyclotron frequency $\w_c$, 
and a random potential $V$. 
We treat each Landau level separately, 
assuming that the magnetic field is strong enough with 
$ |\langle NX|V|N'X' \rangle | \ll \hbar \w_c (N\neq N')$. 

Following Ando\cite{Ando,Ando84,AA}  
we can express the random potential as an assembly of scatterers 
each having a range $d$ as 
\begin{equation} 
   V({\bf r}) = \sum_{i} {V_i \over 2\pi \sigma^2} 
                e^{-|{\bf r}-{\bf r}_i|^2/2\sigma^2 }  \ . 
\end{equation} 
Here the position of the scatterer ${\bf r}_i$ 
is randomly distributed, which 
makes the correlation of the random potential, 
$\langle V({\bf r})V({\bf r}')\rangle \propto (2\pi \sigma^2)^{-1} 
{\rm exp}(-|{\bf r}-{\bf r}'|^2/2\sigma^2)$, Gaussian as well.  
The strength, $V_i$, of the scatterer 
is assumed to be $V_i =\pm V_0$ with an equal probability to 
make the broadened Landau level symmetric. 
In the short-range limit, the potential becomes an assembly of 
$\delta$-functions, 
$V({\bf r})=\sum_{i} V_i \delta ({\bf r}-{\bf r}_i)$. 
The electronic structure is dominated by the 
dimensionless concentration of scatterers, 
$c \equiv 2\pi \ell^2 n_i = n_i/n_L$, where 
$n_i$ is the bare concentration and 
$n_L$ the degeneracy (per unit area) of each Landau level. 

We consider a QH system of size $L$ 
under periodic boundary conditions in $x$ and $y$ 
directions. 
We take system sizes 
as large as $L = (2\pi n_L)^{1/2}\ell = 200\ell \sim 300\ell$, 
which are 
considerably larger than those in previous studies.\cite{aoki83,pook}  
The Hamiltonian is diagonalized by the 
Lanczos algorithm with no re-orthogonalization imposed in 
obtaining the eigenvector.\cite{cullum} 
This algorithm enables us to obtain \lq inner\rq \ eigenvalues 
and eigenvectors accurately with minimal storage requirements 
for large-scale eigenvalue problems. 
The computations of desired eigenvectors have been  carried out 
in two steps. 
First, 
the Hamiltonian is tridiagonalized by means of the Lanczos algorithm, 
and then the tridiagonal matrix is diagonalized in a standard way. 
Second, the eigenvectors are calculated  
>from the series of reproduced Lanczos vectors and 
the eigenvectors of the tridiagonal matrix. 
The size of the tridiagonal matrix is taken to be 
$3N_L \sim 5N_L$ where $N_L$ is the dimension of the Hamiltonian. 
This achieves an accuracy for normalized eigenvectors 
$\psi_{\mu}$ better than 
$\parallel {\cal H}\psi_{\mu} - E_{\mu}\psi_{\mu} \parallel 
= 10^{-10}$ even at the band center, 
where $E_{\mu}$ is the eigenvalue. 

Figure 1 depicts a typical 
spatial behavior of the wave function $|\psi ({\bf r})|^2$ 
that has an eigenvalue $E_{\mu}$ closest to 
the center of $N=0$(a) or $N=1$(b) Landau levels 
for the delta-potentials with the concentration 
$c =10.0$.  
Sparsely distributed spikes are conspicuous in the wave functions 
when shown on a linear scale, 
which is the characteristic feature at a metal-insulator 
transition point.\cite{aoki83,pook} 
This is particularly so for $N=0$, while the $N\geq 1$ 
wave function is more homogeneously distributed.  

We have then obtained 
the multifractal spectrum $f(\alpha )$ 
numerically by the box-counting procedure.\cite{pook,fa}  
Namely, we divide the system 
into $N(\lambda )\equiv 1/\lambda^2$ boxes 
where $\lambda = \ell/L$ is 
the ratio of the box size and the system size. 
The probability $P_j(\lambda )$ of finding an electron in 
the $j$th box $B_j(\lambda )$ is
\begin{equation} 
  P_j (\lambda) \equiv \int_{B_j(\lambda)} d{\bf r} 
  \ |\psi ({\bf r})|^2 \ . 
\end{equation} 
If we define the normalized $q$-th moment of 
the box probability as 
\begin{equation} 
  \mu_j (\lambda ,q) \equiv 
  {[P_j(\lambda )]^q \over \sum_{k=1}^{N(\lambda )}     
  \left[P_{k}(\lambda )\right]^q } 
   \quad ,
\end{equation} 
we can calculate the singularity strength, 
\begin{equation} 
 \alpha (q) = \lim_{\lambda \to 0} 
 { \sum_{j=1}^{N(\lambda )} \mu_j (\lambda ,q) \ln [ P_j (\lambda )] 
   \over  \ln \lambda  } \ , 
\end{equation} 
and the 
multifractal spectrum, 
\begin{equation} 
 f(\alpha (q)) = \lim_{\lambda \to 0} 
 { \sum_{j=1}^{N(\lambda )} \mu_j (\lambda ,q) 
   \ln [ \mu_j (\lambda ,q)] \over \ln \lambda  } \ . 
\end{equation} 
Equations (6) and (7) yield the multifractal spectrum $f(\alpha )$ 
in a parameter $(q)$ representation.\cite{fa} 
The lim$_{\lambda \to 0}$ is taken by reducing the box size to 
the smallest physical size, $\ell$, 
with $\lambda = \ell/L$. 
The generalized dimension $D(q)$ is derived as 
$ D(q) = [f(\alpha (q)) - q\alpha (q)]/(1-q) $. 
Specifically, $D(2)$ gives the fractal dimension 
of the wave function, 
which has been obtained more precisely 
by the direct determination of the box probabilities.\cite{pook}

Figure 2(a) presents the $f(\alpha )$ spectrum of the critical 
wave functions at the center 
of $N=0,1,2$ Landau levels, respectively, where 
a smooth function is obtained 
in a finite region of $\alpha$. 
We have checked that the system size 
is large enough by confirming that $f(\alpha )$'s for two different  
$L=200\ell, 300\ell$ 
agree within statistical errors 
and that we are indeed looking at the critical wave function.\cite{avishai}   
If not, 
the $f(\alpha )$ spectrum would coalesce into a single point 
$(\alpha ,f(\alpha ))=(d,d)$  for extended wave functions, 
or two points, $(\alpha ,f(\alpha ))=(0,0),(\infty ,d)$ 
for localized ones in the limit $L\rightarrow \infty$. 

A salient feature in Fig.2(a) is that 
(i) reflecting the significant 
difference seen in Fig.1, the $f(\alpha)$ spectra does 
vary systematically with $N$, but 
(ii) for all of $N=0,1,2$ $f(\alpha)$ fits very well to 
a parabolic form,\cite{pook} 
\begin{equation} 
  f(\alpha ) = d-{(\alpha -\alpha_0)^2 \over 4(\alpha_0 -d)} \ . 
\end{equation} 
The fact that the critical wave functions at higher Landau levels 
as well as the lowest one possess the parabolic form 
$f(\alpha )$ 
indicates a log-normal distribution 
${\exp}[({\rm ln}P-\alpha_0 {\rm ln}\lambda)^2/4(\alpha_0-d){\rm ln}\lambda]$, 
on the box probability $P$.  
Since the position of the peak $\alpha_0$ ($\alpha$ for $q=0$), 
and the curvature 
are correlated as seen from Eq. (8), $f(\alpha )$ 
is characterized by a {\it single} exponent $\alpha_0(N)$, 
which is a function of $N$. 
The value of $\alpha_0$ decreases from 
$\alpha_0(0) =2.31 \pm 0.02 $ to 
$\alpha_0(1) =2.15 \pm 0.02 $ 
and further down to $\alpha_0(2) = 2.09 \pm 0.01 $. 
This is the first key result of this paper.  
We can visualize this in a typical profile of the 
 $N=1$ wave function on a logarithmic scale in Fig. 3(b). 
The figure shows that the fluctuation in the wave 
fucntion, which is wild on the logarithmic scale, 
indeed becomes similar to 
 the $N=0$ one (Fig. 3(a)) 
 when a linear transformation on logarithmic scale, 
${\rm ln}|\psi | \rightarrow 
 [(\alpha_0(0)-d)/(\alpha_0(1)-d)]{\rm ln}|\psi | 
 + \mbox{const}$, is performed 
 to make the two distributions coincide.   
The approach of $\alpha_0$ to the support of the measure 
(= spatial dimension $d$) thus 
means that the amplitude of wave functions becomes less 
wildly distributed.  

Accordingly, 
the fractal dimension $D(2)=2\alpha (2)-f(\alpha (2))$ of the critical wave functions 
approaches to 
$d=2$ as we go from $N=0 (D(2) = 1.50 \pm 0.06)$ 
to $N=1 (D(2) = 1.73 \pm 0.05)$ 
and $N=2 (D(2) = 1.82 \pm 0.05)$, 
and, consequently, the density autocorrelation, 
$\langle |\psi({\bf r}) \psi({\bf r+R})|^2 \rangle \sim R^{D(2)-d}$, 
becomes longer tailed. 
We have a smaller value for $N=0$ compared with 
that obtained by the lattice model\cite{huckerev} 
or the scaling of the static conductivity,\cite{gammel}  
but the values of $D(2)$ for $N\geq 1$ 
are close to those obtained from the exponents of the 
density-density correlation by Mieck.\cite{Mieck}

If we turn to finite-ranged potentials, $\sigma >0$, 
Figure 2(b) shows the $f(\alpha )$ spectrum for the critical wave function 
for $N=1$ with $\sigma=0.71\ell$ and $c = 3.0$. 
The $f(\alpha )$ spectrum continues to be 
parabolic, but 
the value of $\alpha_0$ significantly increases to $\alpha_0 = 2.32 \pm 0.03 $ 
with decreased $D(2) =1.51\pm 0.07$. 
The respective values of $\alpha_0$ and $D(2)$ 
almost go back to those for $N=0$.  
This is also seen in the profile of the wave function in Fig.1(c). 
The resemblance to the $N=0$ situation 
provides a sign of the recovery to the 
$N=0$ situation.

We can now look at the scaling amplitude 
estimated as $\Lambda_c =[ \pi(\alpha_0-d) ]^{-1} $ 
for $N=1$ as a function of $\beta^2 \equiv 1+(\sigma/\ell)^2$ 
in Fig. 4.  
As the correlation length $\sigma$ is increased, 
the value of $\Lambda_c$ approaches to 
a value that is consistent with 
an argument by Lee {\it et al.}\cite{lee} on the universal conductivity at 
the transition with $\Lambda_c =1/{\rm ln}(1+\sqrt{2})$. 
This recovery of the single-parameter scaling in the 
higher Landau level with long-range potential 
agrees with the finite-size scaling result of 
Huckestein\cite{huckerev}.

In conclusion, 
the multifractal spectrum $f(\alpha )$ 
at criticality does depend on the Landau inded $N$ while 
a universality still exists 
in that its parabolic form 
is preserved, 
and the scaling valid for the lowest 
Landau level is asymptotically recovered as the potential range is increased to $\sim \ell$ for $N=1$.  
One will be able to see how the localization exponent $\nu$ behaves via the 
identification $1/\nu = \alpha_0 +x$\cite{Janssen} 
by calculating the normalization exponent $x$ for 
higher Landau levels. 
The $N$ dependence of the fractality 
will also have an implication on the anomalous 
diffusion coefficient $D(q,\omega) \sim (q/\sqrt{\omega})^{-\eta}$, where 
$\eta=d-D(2)$.

One of the authors (H. A.) wishes to thank Yshai Avishai 
and Hiroshi Imamura for valuable discussions, 
and Wilhelm Brenig for showing a preprint prior to publication.  
This work was supported in part by a Grant-in-Aid from 
the Ministry of Education, Science, and Culture of Japan. 
Calculations were done on the Supercomputer Center, 
Institute of Solid State Physics, 
University of Tokyo.

\newpage
\noindent {\bf Figure captions}\par
\ \\
\par
\noindent {\bf FIG. 1.} 
Squared amplitude of a typical 
wave function on a linear scale at the center 
of each Landau level for the system of size $300\ell$ 
for the $N=0$ Landau level with $\sigma =0$ (top), 
$N=1$ with $\sigma =0$ (middle), and 
$N=1$ with $\sigma =0.71\ell$ (bottom). 
\par 
\ \\
\par 
\noindent {\bf FIG. 2.}  (a) The $f(\alpha)$ spectra (for $|q|\le 4$) 
of the wave functions at the center of the 
$N=0, 1, 2$ 
Landau levels for short-range random potentials ($\sigma =0$). 
The solid circles (squares) represent 
the system size $200\ell (300\ell)$.\par 
(b) The same for a finite-ranged 
random potential, $\sigma =0.71\ell 
(\beta^2 =1.5)$, for $N=1$ Landau level. 
\par 
\ \\
\par
\noindent {\bf FIG. 3.} 
Logarithm of the squared amplitude of a typical 
wave function at the center 
of $N=0$ Landau level with $\sigma =0$ (top) is compared with 
the same for $N=1$ 
with a rescaling of the wave function to make its 
 log-normal distribution the same as for $N=0$ (bottom). 
\par 
\ \\
\par
\noindent {\bf FIG. 4.}  The scaling amplitude 
$\Lambda_c = [\pi (\alpha_0-d)]^{-1}$ for $N=1$  
against $\beta^2 \equiv 1+(\sigma/\ell)^2$.  
The horizontal dashed line indicates 
$\Lambda_c =1/{\rm ln}(1+\sqrt{2})$.\par
\end{document}